\documentclass[12pt]{iopart}
\usepackage{iopams}
\usepackage{graphicx}
\usepackage{dcolumn}
\usepackage{bm}
\usepackage{array}
\usepackage[linktocpage=true]{hyperref}
\usepackage{multirow}
\usepackage{appendix}

\newcommand{\beq}{\begin{equation}}
\newcommand{\eeq}{\end{equation}}

\newcommand{\Eq}[1]{Eq.~(\ref{#1})}
\newcommand{\Equation}[1]{Equation~(\ref{#1})}

\newcommand{\Fig}[1]{Fig.~\ref{#1}}
\newcommand{\FigurePragraph}[1]{Figure~\ref{#1}}
\begin{document}

\title[Single photon absorption   in a 1D waveguide]{Coherent single-photon absorption by single emitters coupled to 1D nanophotonic waveguides}
\author{Yuntian Chen,$^{1,*}$ Martijn Wubs,$^1$ Jesper M\o rk,$^{1}$ and A. Femius Koenderink$^{2}$}
\address{$^1$ DTU Fotonik, Department of Photonics Engineering, \O rsteds Plads, DK- 2800 Kgs. Lyngby,
Denmark.}
\address{$^2$Center for Nanophotonics, FOM Institute for Atomic and Molecular Physics (AMOLF),  Science Park 104, 1098 XG Amsterdam, The Netherlands}
\address{$^*$Email: yche@fotonik.dtu.dk}
\begin{abstract}
We study the dynamics of single-photon absorption by a single emitter coupled to  a one-dimensional waveguide that simultaneously provides channels for spontaneous emission decay and a channel for the input photon. We have developed a time-dependent theory that allows us to specify any input single-photon wavepacket guided by the waveguide as initial condition, and calculate the excitation probability of the emitter,  as well as the time evolution of the transmitted and reflected field. For single-photon wavepackets with a gaussian spectrum and temporal shape, we obtain analytical solutions for the dynamics of absorption, with maximum atomic excitation $\thicksim 40\%$. We furthermore propose a terminated waveguide to aid the single-photon absorption. We find that for an emitter placed at an optimal distance from the termination, the maximum atomic excitation due to an incident single-photon wavepacket can exceed $70\%$. This high value is a direct consequence of the high spontaneous emission $\beta$-factor for emission into the waveguide. Finally, we have also explored whether waveguide dispersion could aid single-photon absorption by pulse shaping. For a gaussian input wavepacket, we find that the absorption efficiency can be improved by  a further $4\%$ by engineering the dispersion. Efficient single-photon absorption by a single emitter has potential applications in quantum communication and quantum computation.
\end{abstract}

\maketitle
\setcounter{tocdepth}{1}
\section{Introduction}
Ultimate control over single light quanta, the emission of single photons, the absorption of single photons and the routing of photons between qubits, is of core interest for quantum information technology and extensively studied in \cite{Raimond2001RevModPhys,GerardPRL1998,PeltonPRL1998,Wilk2002Science,Hennessy2007Nature,Reithmaier2004Nature}.  Ideally, one could use single photons that rarely interact with each other, and the environment as natural messengers of quantum information between nodes, where actual operations take place. Such nodes could then consist of localized atoms or quantum dots that can interact strongly with each other or external stimuli via, e.g., side band coupling, RF fields, or electrical gate signals, in an efficient and controllable way. Thus, it is extremely desirable to map a flying photonic qubit state   onto an atomic qubit   with unit probability \cite{Cirac1995PRL,Cirac1997PRL}. However, such interfacing between an atom qubit and a flying photonic qubit  is challenging, since it requires simultaneously an open photonic system for easy interfacing with freely propagating photons, yet also a high light-matter interaction strength, which is usually associated with the use of a high-Q, closed photonic system surrounding the atomic qubit.
In absence of a high-Q cavity, one can either use a large  ensemble of atoms  to compensate for weak optical transition strengths of single atoms \cite{Julsgaard2004Nauture}, or use highly focused optical beams to excite single atoms or molecules as efficiently as possible \cite{Wrigge2008NaturePhy,Lindlein2007LaserPhysics,Piro2010NatPhy}. Among different mapping techniques, direct absorption of single photons by light emitters is  an attractive option that may be realized using recent advances in the engineering of complex photonic environments \cite{Wrigge2008NaturePhy,vanEnk2004PRA,Sondermann2007APB,Rao2007PRL,
Gerhardt2007PRL,Zumofen2007PRL,Kuhn2006PRL,Chang2007Np,Leuchs2009EPL}. Such efficient single-photon absorption is not only of relevance in quantum optics, but is also of central relevance for detectors, photovoltaics, or optical sensing and microscopy based on absorption or fluorescence \cite{Kukura2009NanoLetter,Gaiduk2010Science}.

\begin{figure}\centering
\includegraphics[scale=0.65]{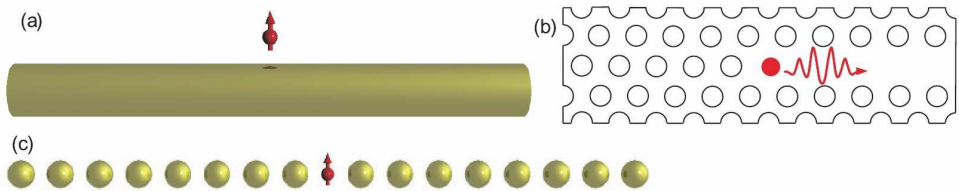}
\caption{\label{sketch1D_WGemitter} Possible experimental realizations of a 1D waveguide
strongly coupled to a single emitter: (a) a propagating  surface plasmon polariton (SPP) mode on a metallic nanowire,
 (b) a guided mode in a photonic crystal waveguide
and (c) a guided plasmonic mode on an infinitely long plasmon particle chain.}
\end{figure}

Recent advances in nanophotonics enable the funneling of almost all the single photons emitted by a single emitter into a single mode \cite{Akimov2007Nature,LundHansen2008PRL}, or directing single photons into narrow beams \cite{Koenderink2009Nanoletters,Curto2010Science,Kosako2010NatPhoton} and even dynamical steering of singe-photon  emission \cite{Chen2010PRB_R}. As sketched  in \Fig{sketch1D_WGemitter}, it has been shown that these one-dimensional (1D) or quasi-1D waveguides can be used to efficiently control the  spontaneous emission \cite{Akimov2007Nature,LundHansen2008PRL,chang053002,Jun2008PRB,Lecamp2007PRL,Chen2010PRB}. Inversely, reciprocity in classical electrodynamics predicts  efficient coupling of single photons with an emitter, provided the incoming single-photon wavepacket is fed through the channel which is associated with a high spontaneous emission rate. Recent theoretical work \cite{Zumofen2007PRL,Chang2007Np,Shen2005,Heugel2010LaserPhysics,Witthaut2010NJP} has hence focused on the interaction of single emitters and single photons guided by 1D waveguides.  However, this work   mainly concerns the reflection and the transmission probability of  single-photon wavepackets of very narrow spectral bandwidth interacting with a single emitter. These models consist of stationary solutions for the interaction of a two-level system with a one dimensional waveguide. In such stationary cases, the photon wavepackets have a quasi-infinite  temporal extent, and hence the atom is essentially in the ground state at all times.  In this case, the atom can simply be treated as a point scatterer \cite{Chang2007Np,Shen2005,Vries1998RevModernPhy}. To optimize the atomic excitation probability, a  time dependent treatment is required to predict which photon wave packets have optimal temporal and spatial profiles for atomic excitation. Originally it was pointed out by Cirac \mbox{\it{et al.}} \cite{Cirac1997PRL,Gorshkov2007PRL}  that a time-reversal symmetric photon wavepacket can be used to efficiently transfer quantum states among distant nodes consisting of a $\Lambda$-type atomic medium. The interesting concept of time-reversal symmetry  was also applied to  two-level atoms, showing that it is indeed possible to perfectly invert an atomic qubit using photon wavepackets that are spatially and temporally the inverse of the photon wavepacket emitted by a qubit through spontaneous emission \cite{Leuchs2009EPL,Rephaeli2010PRA}. However, realizing this prediction requires highly non-trivial pulse shaping, especially at the single-photon level \cite{Kolchin2008PRL}. Thus there is large need for a time-dependent theory that quantifies what atomic excitation efficiencies can be reached using practically achievable single photon wavepacket and photonic structures.

In this paper we develop a time-dependent theory that quantifies what atomic efficiencies  can be reached using practically achievable single photon wavepackets in 1D waveguides. As sketched in \Fig{sketch1D_WGemitter}, we consider a single-photon wavepacket propagating along a 1D waveguide, interacting with a single emitter. We explore the possibility of maximizing the single-photon absorption by the emitter via  engineering the photonic environment of the emitter, and by shaping the pulse  of the input single-photon wavepacket through waveguide dispersion. We study 1D or quasi-1D waveguides, considering all the spontaneous emission  channels, since it is the competition between the pumping channel and the spontaneous emission channel of the emitter that will ultimately determine the absorption efficiency.  We focus on the role of the temporal coherence of the input single-photon wavepacket for exciting the emitter in the 1D waveguide. The formalism can be generalized to treat a three-dimensional (3D) light-scattering problem for examining the role of spatial coherence, which will be addressed in a  future paper.

The paper is organized as follows. In Section \textbf{2}, we formulate the model and derive general solutions for the dynamics.  Section \textbf{3} presents results obtained for  gaussian input single-photon wavepackets propagating along an infinitely long 1D waveguide. We obtain an analytical solution for the dynamics of the atomic excitation. Furthermore, we present a physically transparent model that relates our dynamical model to the stationary reflection and transmission spectrum, and we present a simple model for the time-dependent reflection and transmission. We find that the maximum absorption probability strongly depends on the temporal coherence of the input single-photon wavepacket. In Sec. \textbf{4}, we consider an emitter located near the termination  of a half-truncated 1D waveguide. Near such a termination, the spontaneous emission $\beta$-factor for uni-directional emission can be strongly enhanced. By reciprocity, uni-directional emission with a high beta factor also results in the highest possible atomic excitation probability. Our work thus provides a simple route to optimize photonic waveguide structures for photon-qubit interaction. Finally, we  investigate the possibilities of improving the atomic excitation by pulse chirping, or pulse dispersion to modify the temporal coherence.  Section \textbf{5}  concludes the paper.

\section{Model}
In this section, we will outline our theory for calculating the dynamics of single-photon absorption.
Our model follows a similar procedure employed by Dorner \mbox{\it{et al.}} \cite{DornerPRA2002}  for spontaneous emission and resonance fluorescence of a single emitter coupled to a mirror. In section~\ref{subsec:eqmotion}, we construct a general  model for single-photon absorption valid for  arbitrary (non-absorbing) 3D photonic structures with arbitrarily shaped input wavepackets. In Section~\ref{subsec:dynamics},  we apply the general model to  study the dynamics of single-photon absorption by a single atom coupled to a 1D waveguide \cite{Chang2007Np,Shen2005}. The reflection and transmission coefficients derived in stationary theories can be extracted by evaluating our time-dependent solutions at  times far later than the time interval where the photon wavepacket collides with the atom, and after the emitter has relaxed back into the ground state due to spontaneous emission.


\subsection{Hamiltonian, initial conditions and equations of motion\label{subsec:eqmotion}}
In a finite volume the electromagnetic field can be
decomposed into   discrete modes. In the discrete-mode quantization scheme, the interaction between a single-photon wavepacket and the emitter is modeled by the Hamiltonian
\begin{equation}
{{\hat H}_{\rm tot}} = {{\hat H}_0} + {{\hat H}_I},
\end{equation}
where ${\hat H}_0$ defines the free constituents ${\hat H}_0=\hbar
{\omega _0}|e
\rangle   \langle  e| + \sum\limits_{\lambda} {\hbar \omega_{\lambda} {{\hat
a}^\dag }({\lambda} )\hat a({\lambda} )} $, and where ${\hat H}_I$ defines the
interaction part ${\hat H}_I=\sum\limits_{\lambda} {\hbar
[{g_{\lambda} }{{\hat a} }({\lambda} )|e
\rangle   \langle  g| + g_{\lambda} ^ * {\hat a}^\dag({\lambda})|g \rangle   \langle  e|]}$. Here $|g\rangle $ and $|e\rangle $ represent the atom ground  and excited state respectively, $\omega _0$ denotes the atomic transition frequency, while $\hat a({\lambda} )$ is the annihilation operator for the mode ${\lambda}$. Without loss of generality, we take the emitter position as our real-space origin (emitter located at $\mathbf{r}=0$).
We expand the Schr\"{o}dinger-picture state $|\psi (t) \rangle $ at time $t$
in the basis of all states with one excitation, i.e.,  $|e,0\rangle $,  $|g,1_\lambda\rangle $,  as follows:
\begin{equation}\label{BasesFunction}
|\psi (t) \rangle   = C_0^e(t){e^{ - i{\omega _0}t}}|e,0 \rangle   +
\sum\limits_{\lambda}{ {e^{ - i\omega_{\lambda} t}}C_{{\lambda}}
^g(t)|g{{,1}_{{\lambda}}} \rangle  }.
\end{equation}
By separating the time dependence driven by $H_0$, the coefficients $C_0^e(t)$ and $C_{{\lambda}}
^g(t)$ are essentially solved in the interaction picture.
Since we are interested in the absorption of a single-photon wavepacket, we assume that in the initial state, the atom is unexcited ($C_0^e(0)
= 0$), while the initial wavepacket is described by  $C_{\lambda} ^g(0)=\xi ({\lambda})$.

The equations of motion are
\begin{eqnarray}\label{allequationsofMatrix}
{i\hbar \frac{{dC_0^e(t)}}{{dt}} = \sum\limits_{\lambda}
{C_{\lambda} ^g(t) \langle  e,0|{H_I}|g{{,1}_{\lambda} } \rangle  {e^{ -
i(\omega_{\lambda}  - {\omega _0})t}}} },\label{WW1}\\ {i\hbar
\frac{{dC_{\lambda} ^g(t)}} {{dt}} = C_0^e(t) \langle  g{{,1}_{\lambda}
}|{H_I}|e,0 \rangle  {e^{i(\omega_{\lambda}  - {\omega _0})t}}}.\label{WW2}
\end{eqnarray}
The transition matrix element $\langle g,1_{\lambda}|{H_I}|e,0\rangle  $ can be simplified as $\langle  g,1_{\lambda}|{H_I}|e,0 \rangle  = \hbar g_{\lambda} ^ * $, where $g_{\lambda}$ is the well-known interaction strength of a single emitter with mode $\lambda$ \cite{Loudon2000}. Integrating Eq.~(\ref{WW2}) yields
\beq\label{C0omega} C_{\lambda} ^g(t) = C_{\lambda} ^g(0) -
i\int_0^t {C_0^e(t')g_{\lambda} ^ * {e^{i(\omega_{\lambda}  -
{\omega _0})t'}}dt'}, \eeq
and using this in  Eq.~(\ref{WW1}), one can obtain \beq\label{MasterEquation}
\frac{{dC_0^e(t)}} {{dt}} =  - i\sum\limits_{\lambda}  {C_{\lambda}
^g(0){g_{\bm \lambda} }{e^{ - i(\omega_{\lambda}  - {\omega
_0})t}}} - \sum\limits_{\lambda}  {{g_{\lambda} ^ * {g_{\lambda}
}\int_0^t {C_0^e(t'){e^{ - i(\omega_{\lambda} - {\omega _0})(t -
t')}}dt'} } }. \eeq
\Equation{MasterEquation}  is of central importance in this work:  solving this equation of motion provides the time dependent excitation amplitude of the single emitter, hence quantifying the efficiency of single-photon absorption. The right-hand-side (rhs) of
\Eq{MasterEquation} consists of two terms. The first term accounts for excitation of the atom by the single-photon wavepacket. The second term accounts for de-excitation of the two-level system by emission of a photon. Two facts are immediately obvious: firstly,
\Eq{MasterEquation} reduces to the Weisskopf-Wigner
theory of spontaneous emission when the atom is excited and all the
modes of the field are empty initially \cite{Coldren}. Secondly, the maximum excitation probability of the atom depends on the shape and duration of the incident photon wavepacket. Indeed, the incident photon wavepacket is the only channel for driving the atom, while any excitation is continuously subject to exponential decay due to vacuum fluctuations. In the Weisskopf-Wigner approximation, the second
term of the rhs in Eq.~(\ref{MasterEquation}) is memoryless, and can
be simplified to $\Gamma C_0^e(t)/2$, where $\Gamma
= 2 \pi \sum\limits_{\lambda} {g_{\lambda} ^ * {g_{\lambda} }}
\delta(\omega_{\lambda}- \omega_0)$ is the total spontaneous emission decay rate. By decoupling $C_0^e(t)$ from the time dependence of  $C_{\lambda} ^g(t)$, namely, by combining the two first-order differential equations defined by \Eq{allequationsofMatrix} into a second-order differential equation defined by \Eq{MasterEquation},
one only needs to know  $C_{\lambda} ^g(0)$ in order to solve for $C_0^e(t)$. In \Eq{MasterEquation}, $C_{\lambda} ^g(0)$ represents the mode distribution in the incident single-photon wavepacket. The entire spatial structure of the photonic modes surrounding the emitter is implicitly encoded in the coupling strength $g_{\lambda}$, which can be further expressed as $g_{\lambda}=\frac{\bm \mu \cdot \bm E_{\lambda,1} (\bm r=0)}{\hbar}$, where $\bm \mu$ is the dipole moment of the
emitter, and where $\bm E_{\lambda,1}(\bm r=0)$ is the normalized single-photon field strength for mode $\lambda$, at the position $\bm r=0$ of the emitter.

In the above, we have assumed that the fields belong to a finite quantization volume. In a 1D or quasi-1D waveguide system,  a continuous-mode quantization scheme needs to be adopted. We assume that the single-photon wavepacket supported by the 1D waveguide can be decomposed into cylindrical waves, i.e., $\bm E_{\lambda}(\bm r)=\bm E_{\lambda}(x,y)e^{i(\beta z-\omega t)}$, with the mode label $\lambda=\{m, p, \beta, q\}$, where $\beta$ is the component of the wave vector along the $z$ axis, $q$ represents the magnitude of the wave vector perpendicular to the $z$ axis, $m$ is the angular momentum, and the index $p$ is used to distinguish between two degenerate polarization modes for given $m$, $\beta$, and $q$. For more details about the normalization of these continuous modes and the substitution of summations over discrete modes by integration over continuous modes  see Appendix A. We also assume that the incoming single-photon wavepacket has a very narrow bandwidth, on the order of $10^9$ rad/s, since its temporal duration is comparable to the SE lifetime. Such a narrow bandwidth with respect to the atomic transition frequency $\omega_0=10^{15}$ rad/s ensures the the validity of the linear dispersion relation, i.e., $\beta_\omega-\beta_0=(\omega  - {\omega _0})/{v_g}$, where $v_g$ is the group velocity of the propagating mode. From \Eq{SUM} and the Weisskopf-Wigner approximation, one
obtains, \beq\label{ReducedMasterEquation}
\fl \frac{{dC_0^e(t)}} {{dt}} =  - \frac{ie^{  i\omega_0 t}}{\sqrt{v_g} \hbar} \bm \mu
\cdot [ \int d\omega\sum\limits_{\kappa } {\chi_{\omega,\kappa }^g(0){\bm E_{\omega,\kappa,1,con } (\bm r=0)}{e^{ - i\omega
t}}} ]- \Gamma C_0^e(t)/2,\eeq where $\chi_{\omega,\kappa }^g(t)=\frac{C_{\omega,\kappa }
^g(t)}{\sqrt{\triangle\beta v_g}}$, and $\kappa$ is given by $\kappa =\{m, p, q\}$.
\Equation{ReducedMasterEquation} can be used to calculate the excitation probability of the single atom as it is illuminated by a single-photon wavepacket in a complex photonic environment, e.g., in a photonic cavity or near an optical antenna.


\subsection{Dynamics  of absorption, reflection, and transmission in  1D and   quasi-1D waveguides\label{subsec:dynamics}}
In this section we consider the dynamic response of both the emitter and the photon field, taking as initial condition a single-photon wavepacket incident from the -z direction, traveling towards the emitter at $z=0$. The transverse distribution of the waveguide mode is independent of z, and the longitudinal wave number of the modes $k$ is assumed to obey the linear dispersion relation $k-k_0=(\omega-\omega_0)/v_g$  in the frequency range  of interest. We study the dynamics of coherent single-photon absorption with special focus on the dependence of absorption on the properties of the input single-photon wavepacket. The dependent variable time $t$ has a one-to-one correspondence with the position of the peak of the single-photon wavepacket. Initially, the peak of the pulse is at the position $z=-Z_0$, and the interaction between the single-photon wavepacket with the emitter reaches its maximum at $t=Z_0/v_g$. At approximately this time, the emitter reaches its maximum excitation probability. Meanwhile, the emitter also loses atomic excitation probability due to the fact that it decays via spontaneous emission in both the forward and backward direction. The emitted light  can interfere with the incident beam,  which results in extinction and pulse reshaping in the transmitted channel.  The relevant timescale for de-excitation by  spontaneous emission is the lifetime  $1/\Gamma$, where $\Gamma$ is the spontaneous emission decay rate of the emitter. The single-photon wavepacket will  be  redistributed  into a (re-shaped) reflected and transmitted pulses and   the emitter will  relax  to the ground state  in several lifetimes after  attaining the maximum excitation probability.

In any 1D waveguide, the continuous-mode variable can  equivalently be taken as the wavevector $k$ or the frequency $\omega_k$ due to the linear dispersion approximation. The wavevector $k$ is chosen here.  In such infinitely long single mode waveguide, the spontaneous emission divides equally between the forward and backward propagating channel. If we assume that the incident photon wavepacket is a single packet offered from just one direction, this means that the spontaneous emission $\beta_0$-factor    for the pumping channel is at most $50\%$.
For any given incident photon wavepacket $C_k ^g(0)$, it is straightforward to calculate the dynamic  atomic excitation $C_0^e(t)$ in the 1D waveguide, which is given by,
\beq\label{MasterEquation1D} \frac{{dC_0^e(t)}} {{dt}} =  -
i\int\limits_{-\infty}^\infty  {dk C_k ^g(0){g_k }{e^{ -
i(\omega_k  - {\omega _0})t}}}  - \Gamma C_0^e(t)/2.\eeq In
our case, the coupling strength ${g_k }$ is assumed to be frequency-independent due to the narrow bandwidth of the input single-photon
pulse and given as ${g_k}= \sqrt {\frac{{\Gamma }} {{4\pi }}v_g}$  \cite{Chang2007Np}.

Once the atomic excitation $C_0 ^e(t)$ is known, one can  calculate the dynamics of $C_k ^g(t)$  to obtain the reflected and transmitted wavepacket. According to
Eq.~(\ref{C0omega}), the total probability amplitude of  the forward
propagating wavepacket is a coherent superposition of the initial incident
photon wavepacket and the emitted photon, and is given by
\beq\label{transmission} C_{k,+} ^g(t) = C_k ^g(0) -
i\int_0^t {C_0^e(t')g_k ^ * {e^{i(\omega_k  - {\omega _0})t'}}dt'}.
\eeq For the backward direction,  there is no incident term ($C_{k,-} ^g(
0)=0$). Hence, the corresponding probability amplitude is \beq\label{reflection}
C_{k,-} ^g(t) = - i\int_0^t {C_0^e(t')g_k ^ * {e^{i(\omega_k
- {\omega _0})t'}}dt'}. \eeq
In most realistic cases, an emitter coupled to a 1D waveguide will still have a residual coupling to free space modes that are not guided by the wire.  For such a quasi-1D waveguide, the spontaneous emission $\beta$-factor into the forward plus backward propagating waveguide modes will have a value  less than $100\%$. Also, a break in symmetry in the geometry might imply that  emission into forward and  backward propagating waveguide modes is unbalanced. In these cases, where the pumping channel funneling  the  incident single-photon wavepacket has  a spontaneous emission  $\beta$-factor of $\beta_0$,
\Eq{MasterEquation1D} becomes, \beq\label{MasterEquation1DBeta}
\frac{{dC_0^e(t)}} {{dt}} =  - i\int\limits_{-\infty}^\infty  {dk
C_k ^g(0){\sqrt {\frac{{{\beta_0 \Gamma }}} {{2\pi }}v_g
}}{e^{ - i(\omega_k - {\omega _0})t}}}  - \Gamma
C_0^e(t)/2.\eeq
It should be remarked that the first term of the rhs  in \Eq{MasterEquation1DBeta} is simply the pulse shape in absence of light-matter coupling, assuming that the coupling strength $g_k$ is a constant. In the next section we will study specific wavepackets and make use of the equation of motion derived here.
\section{Gaussian pulse excitation of one-dimensional waveguide}
First we consider  gaussian input single-photon wavepackets
defined as $ C_k ^g(0) ={[\frac{2} {{\pi {\Delta
^2}}}]^{\frac{1}{4}}}{e^{[ i(k-{k _0} ){Z_0} - \frac{{{{({k _0} -
k )}^2}}} {{{\Delta ^2}}}]}}$, where $-Z_0$ is the position at which
the peak of the pulse passes at $t=0$. The relation $\int_{\infty}^{\infty}| C_k ^g(t = 0)|^2dk=1$ ensures that there is only a single-photon in the wavepacket. With the linear dispersion relation and the substitutions of $t_0=\frac{Z_0}{v_g}$ and $\Omega=\Delta v_g$, one can obtain an analytical solution for the dynamics of single-photon absorption, i.e., for the excitation amplitude of the two level system
\beq\label{DynamicalSolution}
C_0^e(t)=s\frac{1}{2}\sqrt{\frac{\pi }{a}}e^{\frac{b^2}{a} - c}[\mbox{\rm erf}(\sqrt a t+\frac{b}{\sqrt a})-\mbox{\rm erf}(\frac{b}{\sqrt a})]e^{- \frac{\Gamma}  {2}t},
\eeq
where $s = - i\sqrt {\frac{{{\Gamma}}}{{4\pi }}}
{[2\pi {\Omega ^2}]^{\frac{1} {4}}}$, $a = \frac{{{\Omega ^2}}}{4}$,
$b = - ({\Gamma } - {\Omega ^2}{t_0})/4$,  $c = {\Omega
^2}t_0^2/4$, and $\mbox{\rm erf}(x)$ is the error function, defined as
$\mbox{\rm erf}(x)=\frac{2}{\sqrt{\pi}}\int_0^{x}e^{-t^2}dt$. The result $C_0^e(t)$ obtained from the time-dependent theory \cite{DornerPRA2002,RistPRA2008} contains much richer information than can be obtained from stationary theory, due to the fact that the initial conditions can not be included in the Fourier transform descriptions of light scattering by emitters \cite{Shen2005,Chang2007Np,DornerPRA2002,RistPRA2008}.  However, we would like to point out that an alternative route to solve the problem that includes the initial condition is the Laplace transform method developed by one of us (See  Wubs \mbox{\it{ et al.}} \cite{Wubs2004PRA}).
We find that the two methods are equivalent and yield the same result for $\left|C_0^e(t)\right|^2$. The derivation of $|C_0^e(t)|^2$ based on the Laplace transform method is briefly outlined in Appendix B.
%
%

\begin{figure}\centering
\includegraphics[scale=0.55]{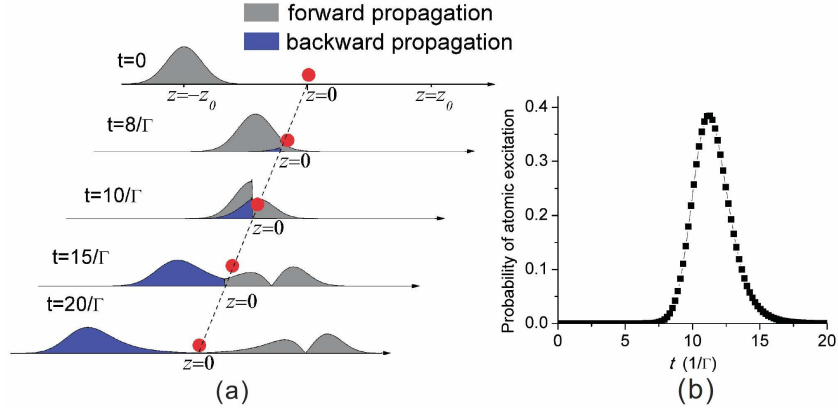}
\caption{\label{SinglePhotonAbsorptionFig2} (a) Snapshots of the amplitudes of the
transmitted and reflected wavepackets as distributed along the spatial axis at different instances of time, (b) the time dependence of the atomic excitation. At $t=0$, the peak of the gaussian input single-photon wavepacket is localized at $z=-Z_0=-10 v_g /\Gamma$, with FWHM
equal to $\Gamma$.  The red dot indicates the fixed position of the emitter.}
\end{figure}

\FigurePragraph{SinglePhotonAbsorptionFig2}(a) shows the spatial variations of the
forward and backward propagating wavepackets, which are obtained by
Fourier-transforming $C_{k,+} ^g(t) $ and  $C_{k,-}
^g(t) $ with respect to $k$.  Initially  the incident single-photon wavepacket is  far  away ($z<-Z_0$) from the emitter. At this instance there is negligible interaction. As the leading edge of the pulse reaches the emitter, the interaction is effectively switched on. From this time onwards, the emitter gains excitation amplitude, and starts to emit an outgoing wavepacket into both the forward and backward directions. Before the emitter reaches its maximum excitation shown in \Fig{SinglePhotonAbsorptionFig2}(b), one notices that the forward-propagating wavepacket experiences a sharp drop at the position of the emitter, while the backward-propagating wavepacket experiences a steady increase of its magnitude at $z=0$. \FigurePragraph{SinglePhotonAbsorptionFig2}(b) shows the time dependence of the corresponding atomic excitation, which will be discussed in sub-section (\emph{3.2}). The sharp drop in the forward propagating wavepacket is due to the transfer of energy from light to the emitter, the rate of which is larger than the SE decay rate of the emitter itself, given our assumption on the coupling strength $g_k$, set by the mode profile and $v_g$. In contrast to the forward propagating packet, the backward propagating wavepacket obtained purely from  emission has an increasing magnitude. After reaching the maximum excitation, the relaxing emitter leads to decreasing magnitudes for both
forward and backward propagating wavepackets. Moreover, the forward-propagating wavepacket is seen to be strongly reshaped due to  interference of the incident packet with the emitted light. The resulting minimum in the forward propagating wavepacket will be further discussed in the stationary limit in sub-section (\emph{3.1}). One also notes that the trailing edges of the two wavepackets are longer than the leading edges. The widths of the leading edges are determined by the incoming single-photon wavepacket, while the trailing tails are essentially determined by the lifetime of the excited state of the emitter.


\subsection{Stationary limits of the single-photon absorption in 1D waveguide}
We now examine our theory in the stationary limit, to connect our dynamic results to earlier work by Shen and Fan \cite{Shen2005}.  The stationary limit essentially corresponds to  single-photon pulses with  large temporal widths and  low amplitudes, implying also that the emitter is always in the ground state. From our dynamic theory, we can extract transmission and reflection amplitudes in the limit of long excitation pulses, and at times far later than the time window in which interaction with the atom has taken place. Using a Fourier transform, we extract frequency dependent reflection and transmission amplitudes. We compare these frequency dependent coefficients, which we refer to as the stationary limit of our dynamic theory,  with the frequency-dependent coefficients obtained in purely stationary theory by Shen and Fan \cite{Shen2005}. In the stationary limit, in which  the emitter has relaxed into its ground state, the forward- and backward-propagating wavepackets are essentially the transmitted and reflected light, similar to light scattering by an impurity in a 1D waveguide. If we take a snapshot long after the interaction time (we use $t_s =100 /\Gamma$, but our results are independent of this choice), we find instantaneous reflection and transmission amplitudes in $k$-space. By taking advantage of the linear dispersion relation, we can convert these $k$-space instantaneous  amplitudes into spectral information. The transmission and reflection spectra can be extracted as
\begin{eqnarray}\label{TransRefle000}
T({\omega}) =  \left|C_{\omega_k,+} ^g(t_s)/C_{\omega_k} ^g(0)\right|^2 ,\\
R({\omega}) =  \left|C_{\omega_k,-}^g(t_s)/C_{\omega_k} ^g(0)\right|^2 .\label{RcoefficientOmega}
\end{eqnarray}
At first sight, the coexistence of the time and frequency dependence in the transmission and reflection spectra may appear odd. However, for times $t_s$ much later than the interaction time, the time $t_s$ is essentially equivalent to a spatial variable $z=v_g t_s$. Hence, the transmission and reflection spectrum   can be interpreted as frequency signals monitored at a certain position.
%
\begin{figure}\centering
\includegraphics[scale=0.7]{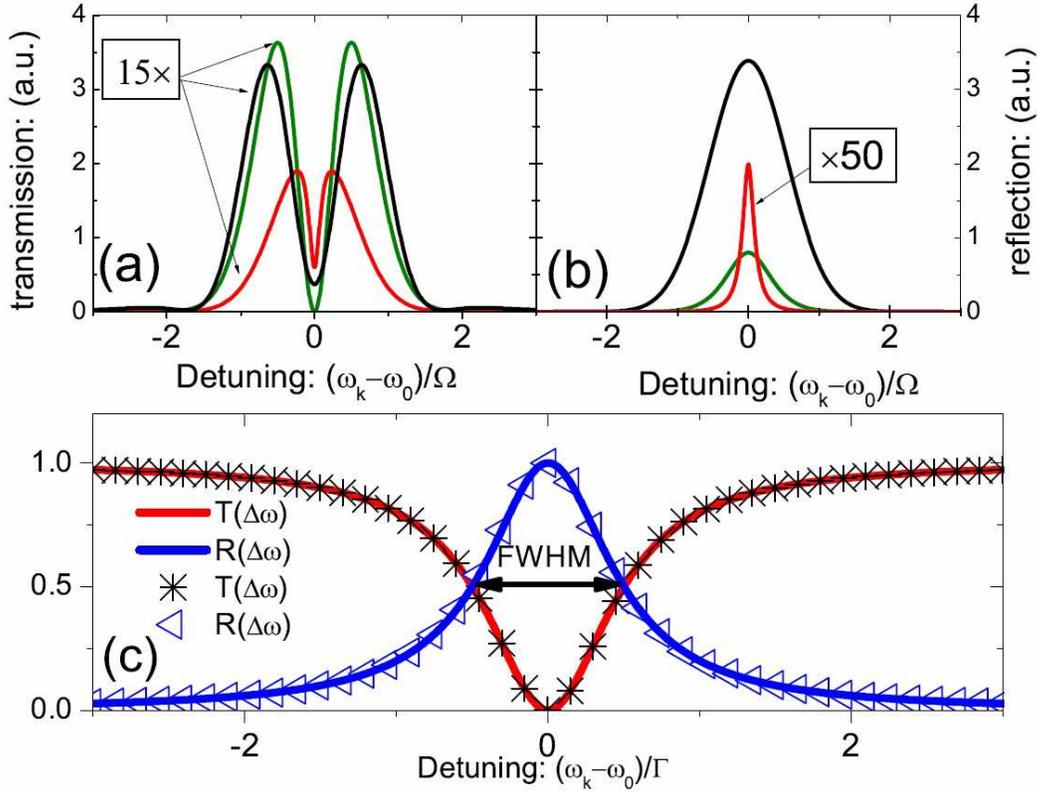}
\caption{\label{CompareFan} (a) Transmitted spectral intensity and (b) reflected
spectral intensity of a gaussian incident single-photon wavepacket mediated by a
single emitter. Plotted results are for different ratios of pulse bandwidth $\Omega$ to decay rate $\Gamma$. Note that the x-axis represents  frequency in units of the pulse bandwidth in (a), (b) but in units of decay rate in (c). In (a), (b) we find that as the ratio of bandwidth to decay rate is decreased from large ($5\Gamma$, red),  to small ($0.2\Gamma$, black), the reflected intensity appears to increase strongly. The transmitted intensity shows a pronounced minimum at zero detuning. This minimum is deepest, reaching $T=0$,  when the incident pulse width exactly matches the lifetime of the emitter (dark green, $\Omega = \Gamma$). (c) shows the transmission and reflection spectra,  obtained by normalizing the transmitted and reflected intensity as plotted in (a), (b) to the intensity spectrum of the incident wavepacket. The stars and triangles correspond to the results extracted from our time dependent formalism, via Eq. (12). The solid lines show the stationary calculations.}
\end{figure}

\FigurePragraph{CompareFan} shows the spectral intensities of the transmitted wavepacket and the reflected wavepacket, as well as the transmission and reflection spectra in the stationary limit, at $t_s$, i.e., long after the interaction of the emitter with the incident wavepacket. \FigurePragraph{CompareFan}(a) and (b) show that the spectral intensity of  transmitted light has a dip at the resonance frequency, while the spectral intensity of the reflected light resembles the original pulse shape. The dip  corresponds to the minimum in the forward propagating  wavepacket in \Fig{SinglePhotonAbsorptionFig2}(a) due to the resonant interaction with the emitter, and the dip magnitude  depends on  the bandwidth of the incident wavepacket. We also note that narrower bandwidth will yield more reflection of light. \FigurePragraph{CompareFan}(c) shows the transmission and reflection spectra in the stationary limit, by normalizing the reflected and transmitted pulse spectra to the spectrum of the incident wavepacket. We now compare the transmission and reflection spectra from our limiting procedure with stationary solutions obtained by solving for the eigenstates of the system as reported by Shen and Fan \cite{Shen2005}. Such stationary behavior of the transmission and reflection can be well modeled by solving for the eigenstates of the system. For a given detuning $\delta=\omega_k-\omega_0$, the corresponding reflection and transmission spectrum can be described as follows  \cite{Shen2005},
\begin{eqnarray}\label{TransRefleShen}
T(\delta) = \left|\frac{\delta}{i\Gamma/2+\delta}\right|^2,\\ \nonumber
R(\delta) =\left|-\frac{i\Gamma/2}{i\Gamma/2+\delta}\right|^2.\label{RcoefficientTransRefleShen}
\end{eqnarray}
The transmission  and reflection spectra  that our time-dependent theory predicts using  a  gaussian input wavepacket with narrow bandwidth agree well with Shen's stationary model defined by \Eq{TransRefleShen}, as shown in \Fig{CompareFan} (c). We also like to point out that the  transmission  and reflection spectra for narrow bandwidth pulse in our stationary limit are independent of $\Omega$. This independence, which is rigorous only in the limit of zero bandwidth, is confirmed by the identical  transmission  and reflection spectra for the  three pulses in \Fig{CompareFan} (a) and (b).  Based on the stationary model, the full width half maximum (FWHM) of transmission and reflection equals the total  decay rate of the emitter, i.e., FWHM=$\Gamma$.
\subsection{Time dependence of the transmittance and reflectance of the single-photon wavepacket in 1D waveguide}
The stationary model accounts for the transmission and reflection at times long after the interaction for  any given frequency distribution. Indeed, a good approximation of the  transmitted and reflected wavepackets are obtained simply by multiplication of the incident spectrum, with transmission and reflection coefficients combined with the right phase factor. However, the stationary model does not account for the atomic excitation probability of the emitter during the interaction process.  This dynamic information is an essential result of our model, that we now proceed to discuss. Also,  the model allows to obtain the time dependent transmission and reflection probabilities at the time of interaction, instead of being limited to times much later than the interaction interval. We define the time dependent transmission (reflection) probabilities by including all amplitude traveling to the right (left) as follows:
\begin{eqnarray}\label{TransRefleProbab}
\texttt{T}(t) = \int\limits_{-\infty}^\infty  {dk |C_{k,+} ^g(t)|^2},\label{Tcoefficient}\\
\texttt{R}(t) = \int\limits_{-\infty}^\infty  {dk |C_{k,-}
^g(t)|^2}.\label{RcoefficientT}
\end{eqnarray}
It is important to realize here, that these coefficients are defined not via spatially separating the amplitude present to the left/right of the atom, but strictly by separating forward and backward direction by wave vector. As opposed to the reflection and transmission coefficients in the stationary limit, these instantaneous coefficients have no frequency content. Instead they allow us to monitor the distribution of excitation between the atom, and the forward and backward emission channel as a function of time. The frequency dependent transmission and reflection reported in \Fig{CompareFan} can be obtained by evaluating T(t) and R(t) for large times (e.g., $t > 100 /\Gamma$),  for many narrow-band initial conditions centered at different frequencies.
\begin{figure}\centering
\includegraphics[scale=0.55]{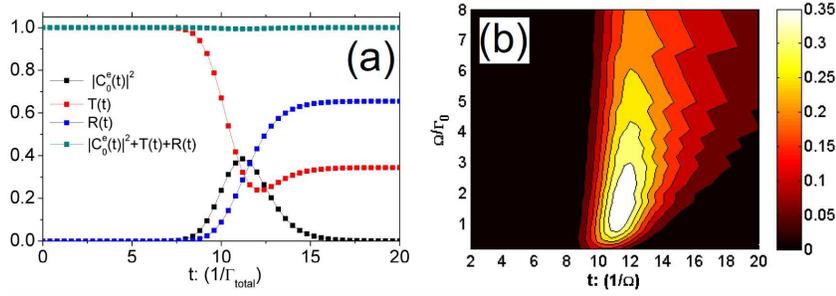}
\caption{\label{Fig4MaximumExication} (a) Time dependence of the
transmission and reflection probability of the incident single
photon wavepacket and the atomic excitation. (b) Contour plot of the
atomic excitation probability versus time and  bandwidth of the incoming
gaussian wavepacket.}
\end{figure}

In \Fig{Fig4MaximumExication}, we consider  the time-dependent probabilities for the atomic excitation, transmission and reflection. For an incoming single-photon wavepacket with FWHM equal to $\Gamma$, we find that the atomic excitation increases as a function of time  as the peak of the pulse approaches . After $t =10/\Gamma$, the atomic excitation decays with a time constant give by the lifetime of the emitter ($1/\Gamma$), after achieving  a maximum value of atomic excitation close to $40\%$, see \Fig{Fig4MaximumExication} (a). The reflection  builds up with time due to the continuous re-emission of light from the emitter.  In contrast, the forward packet has a non-monotonic, though generally decreasing behavior. The kink in $T(t)$ is due to interference of the incident packet and re-emitted light. In contrast to the prediction of the stationary model, the transmitted wavepacket does not vanish, although the incident wavepacket is tuned exactly to resonance. The imperfect reflection as compared to the perfect reflection in the stationary limit is due to the fact that optimum atomic excitation requires a finite pulse length, or equivalently finite bandwidth of the incident packet, whereas perfect reflection only occurs at a  single frequency. The impact of the finite bandwidth of the incoming wavepacket on the maximum value of the atomic excitation is also evident in \Fig{Fig4MaximumExication} (b), which shows the atomic excitation probability as function of time for different bandwidths of the incoming wavepacket. The graph clearly shows that there is a range of optimum bandwidths, that lead  to  atomic excitation probability  with a  maximum value of around  $40\%$. This range of optimum bandwidths is comparable to $\Gamma$, corresponding to incident photon wavepackets that have a duration comparable to the spontaneous emission decay time. The  $40\%$ atomic excitation is surprisingly high given that reciprocity sets the fundamental limit for excitation from just one direction in the wave guide to be  $50\%$. The value  of  $40\%$ could be pushed closer to the limit of 50\% by not using Gaussian pulses, but rather the inverting pulses proposed by Stobi\'nska \rm{et al} \cite{Cirac1997PRL,Gorshkov2007PRL,Leuchs2009EPL,Rephaeli2010PRA}. However, exceeding the limit of  $50\%$ will invariably require illuminating the emitter from two sides with a proper phase relation between the two input pulses. In the following we show that instead of increasing the complexity of illumination by using two pulses, it is also possible to further increase the excitation probability above  $50\%$ by engineering the photonic environment of the emitter to have broken symmetry.

%
%


\section{Gaussian input on a semi-infinite one-dimensional waveguide}
For an infinitely long 1D waveguide, the
atomic excitation probability is prevented from reaching unity by the fact that  the emitter can decay equally
into two directions, namely the forward and backward directions.
Essentially, the coupling efficiency of the emitter with the one-sided pumping channel,  or conversely the spontaneous emission $\beta$-factor in just one waveguide direction, is at best $50\%$.
In order to obtain  higher atomic excitation probability,
all the channels into which the emitter decays should be suppressed,
except  for the one optical input/output channel through which the incident photon wavepacket is sent in. In this section, we consider a situation  where the single-photon wavepacket is launched into a semi-infinite
1D waveguide. By placing the emitter at an optimized distance from the waveguide termination, we expect that the coupling of the emitter to the input waveguide mode is optimized.

\begin{figure}\centering
\includegraphics[scale=0.55]{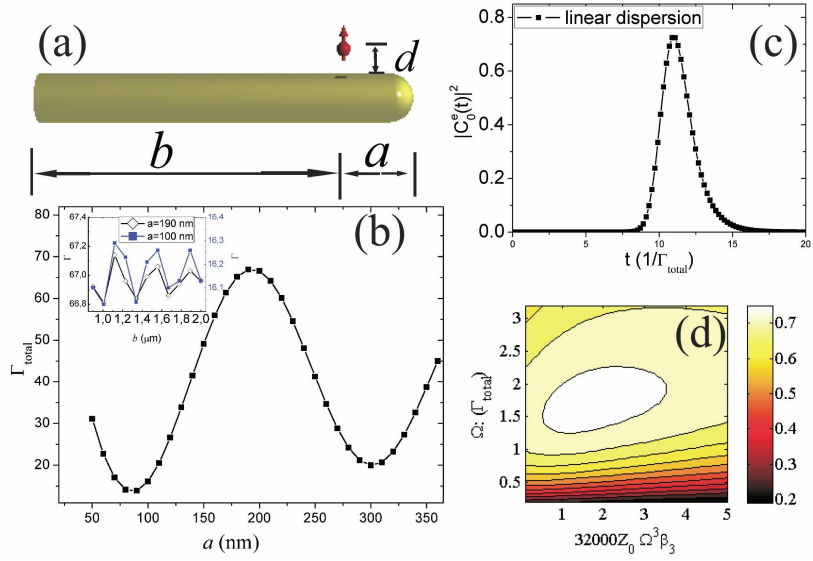}
\caption{\label{Terminatednanowire} (a) Sketch of the terminated
metallic nanowire coupled to a singe emitter. (b) Position
dependence of the total decay rate of the emitter coupled to the
half truncated metallic nanowire. Here $a$ denotes the distance of the
emitter to the rod end on the right. The 20 nm radius silver
nanowire ($\varepsilon_{Ag}$=-50+0.6j) is embedded in the host
material with index $n$=1.414. The inset in (b) shows the total
decay rate as a function of length of the modeling domain, which
does not give any significant impact on $\Gamma$, more details see Ref. \cite{Chen2010PRB}. (c) Time
dependence of the atomic excitation for a terminated metallic
nanowire with spontaneous emission $\beta$-factor of $91\%$. (d) Contour plot of the
maximum of the atomic excitation versus bandwidth  and $\beta_3$ for
a gaussian input wavepacket in a terminated nanowire with
$\beta_2=0$.}
\end{figure}
\subsection{Enhanced spontaneous emission and the spontaneous emission $\beta$-factor by a terminated metallic nanowire}
Specifically, we study a terminated metallic nanowire  coupled to an emitter, as shown in \Fig{Terminatednanowire}(a). On the right-hand side, the metallic nanowire is terminated with a spherical endcap of radius $R$ (equal to the radius of the nanowire). On the left-hand side, the nanowire is infinitely long. As shown in \Fig{Terminatednanowire} (a), $a$ is the distance of the emitter to the right rod end, and $d$ is the distance to the wire edge. \FigurePragraph{Terminatednanowire}(b) shows the variation of the
decay rate as a function of $a$ for $d$=10 nm, and  an emission wavelength of 1000 nm. We obtained  numerical results using finite-element method (FEM) calculations, as explained in our previous work \cite{Chen2010PRB}. Such FEM modeling is rather flexible and can handle complex photonic structures \cite{Chen2010OE} compatible with current lithographic fabrication technology. For an infinitely long 20 nm radius silver nanowire, there are only two guided plasmonic modes, i.e., a forward and a backward propagating guided mode with propagation constants $\pm k$ and corresponding wavelength $\lambda_{\rm eff}=2\pi/k=$436 nm. The decay rate into the forward and backward propagating plasmon mode is 33.8, when normalized to the decay rate of the emitter in vacuum. Accordingly, the decay rate into one direction is $\gamma_{pl,0}\sim$ 17. The total decay rate coupled to an infinitely long nanowire, including the decay into radiative modes in the continuum, and direct quenching, is $\Gamma$=38.27.
Hence the beta-factor into the pumping channel is approximately $44\%$. For a terminated metallic nanowire, the total decay rate shows a clear interference pattern as a function of the distance of the emitter to the termination. This pattern is exactly the 1D analog of of Drexhage's observation that the lifetime of a molecule in front of a mirror oscillates as a function of mirror-emitter separation \cite{Drexhage1970}.  Compared to the infinitely long wire with $\Gamma$=38.27, we can enhance or inhibit $\Gamma$ by a large amount, up to $\thicksim 80\%$, by tuning the source-termination distance $a$. Importantly, by integrating the power flux in the transverse plane of the plasmon nanowire for $a$ = 190 nm, we find that  $91\%$ of the spontaneous emission is coupled into the guided mode that exits the plasmon wire on the left hand side. We hence also expect up to $91\%$ excitation efficiency when offering a single photon wavepacket through the waveguide.

The mechanism responsible for the enhancement is that the waveguide termination essentially functions as a mirror, which reflects the right propagating plasmonic mode to the left. Similar to image dipole theory for Drexhage's experiment, which predicts a two-fold enhancement of the spontaneous emission decay rate for a dipole emitter in front of and perpendicular to a mirror, the waveguide termination also acts as a mirror, which  yields an enhancement factor of 2 of the spontaneous emission into the plasmonic modes. Consequently, the emission rate  into just the left propagating plasmonic mode is approximately enhanced by a factor of 4  since the two-fold enhancement is distributed over just half the channels. Compared to the infinitely long nanowire, such termination with a proper distance to the emitter gives rise to a spontaneous emission $\beta$-factor up to  $91\%$ for the pump channel.
\subsection{Single photon wavepacket propagation along the terminated metallic nanowire}
Having quantified the spontaneous emission decay rate enhancement at the wire termination using FEM modeling,  we now proceed to model the probability of absorption of a single photon wavepacket. To this end we first need to estimate the shape of the wavepacket including the reflected part, for which we need to know the complex reflection coefficient at the wire end. We can estimate the complex reflection coefficient from the calculated decay rate. Indeed, if the reflection coefficient at the rod end is $re^{j\theta}$, the decay rate of the emitter influenced by the rod end can be approximated as $\frac{\gamma_{pl}}{\gamma_{pl,0}}\propto|E_t|^2/|E_0|^2=|E_0(1+re^{j\theta}e^{j2k_0 a})|^2/|E_0|^2$, where $E_0$ is the electric field for the plasmonic mode without reflections, $E_t$ is the total field including the left-hand propagating plasmonic mode and the one reflected from the right. From \Fig{Terminatednanowire} (b), we extract the reflection parameter $|1+re^{j\theta}e^{j2k_0 a} |\sim$ 1.9, since  $\frac{\gamma_{pl}}{\gamma_{pl,0}}\sim 3.6$ according to our FEM calculations.

We now use the reflection coefficient to construct the full wave packet driving the atom, assuming that we initially launch a gaussian wavepacket $\phi_i(z,t) =(2\pi\Delta^2)^{1/4}e^{[i(k_0 z-\omega_0 t)]}e^{[-\frac{\Delta^2(z-v_g t+Z_0)^2}{4}]}$  from $z=-Z_0$ at $t=0$ into the waveguide. The backward propagating wavepacket due to the reflection can be written as
\beq\label{ReflectedField}
\phi_r(Z,t) =r e^{j\theta}(2\pi\Delta^2)^{1/4}e^{ik_0 a}e^{(i(k_0 Z+\omega_0 t))}e^{[-\frac{\Delta^2(a-v_g t+Z_0-Z)^2}{4}]},
\eeq
where $Z=z-a$. Hence the total field given by $\phi_t(z,t)=[\phi_i(z,t)+\phi_r(Z,t)]$ at  $t=0$ is
\begin{eqnarray}\label{totalField}
\phi_t(z,0)=& (2\pi\Delta^2)^{1/4}e^{(i(-\omega_0 t))}e^{[-\frac{\Delta^2(z+Z_0)^2}{4}]}  \\ \nonumber
&+  r e^{j\theta}(2\pi\Delta^2)^{1/4}e^{i2k_0 a}e^{-i\omega_0 t}e^{[-\frac{\Delta^2(z+2a+Z_0)^2}{4}]}.
\end{eqnarray}
We simplify this expression using one assumption, namely that the amplitude envelope factor $e^{[-\frac{\Delta^2(z+2a+Z_0)^2}{4}]}$ in the second term can be simplified to be equal to the amplitude envelope factor $e^{[-\frac{\Delta^2(z+Z_0)^2}{4}]}$ of the first term. The rationale for this approximation is that we consider single-photon wavepackets with a bandwidth of a few gigahertz, corresponding to an envelope length of a few tens of centimeters.  The offset $2a$ in the envelope factor of the second term, which is 2$a$=380 nm, is  negligible compared to  the entire envelope length. In other words, the femtosecond travel time between the emitter and the wire termination is far below the nanosecond temporal length of the pulse envelope. As a consequence, we can take the envelope of the reflected pulse to be identical to that of the incident wavepacket. In this approximation, the total field that appears in the excitation term  is
\beq\label{totoalfieldAppr0}\phi_t(z,0)\simeq\phi_i(z,0)[1+r e^{j(\theta+2k_0 a)}].
\eeq
Exactly as in the case of a symmetric waveguide, the dynamics of the atomic excitation in \Eq{MasterEquation} is set by the field amplitude $\phi_t(k,0)$, i.e., the Fourier transform of $\phi_t(z,0)$, and the coupling strength $g_k$ for the corresponding mode. Due to the reflection from the rod end, the field is enhanced by a factor of $[1+r e^{j(\theta+2k_0 a)}]$, as evident in \Eq{totoalfieldAppr0}.  Not only the field, also the coupling strength $g_k$ is affected by the presence of the wire termination, since the decay rate into the plasmonic channel, i.e., $\Gamma_{pl}=\beta_0\Gamma=2\pi[1+r e^{j(\theta+2k_0 a)}]^2 g_k^2/v_g$, again contains the reflection at the wire termination.
The overall effect of the strong reflection at the wire end on the integral kernel in the excitation term is given by
\begin{eqnarray}\label{excitationTerm}
C_{k,t} ^g(0){g_k }&=[1+r e^{j(\theta+2k_0 a)}]\phi_i(k,0)\sqrt{\beta_0\Gamma v_g/(2\pi[1+r e^{j(\theta+2k_0 a)}]^2)}\\ \nonumber
&=\phi_i(k,0)\sqrt{\beta_0\Gamma v_g/(2\pi)},\end{eqnarray}
where $\phi_i(k,0)$ is the Fourier transform of initially incident wavepacket $\phi_i(z,0)$. \Equation{excitationTerm} shows that the factor $[1+r e^{j(\theta+2k_0 a)}]$ cancels out in the excitation term.  Therefore, the equation defined in \Eq{MasterEquation1DBeta} has properly taken the reflection coefficient as well as the reflection phase into account, except that we assume the   envelope  to be constant over the spatial range from the emitter to the rod end, which is a valid  approximation as  discussed.

\subsection{Enhanced single-photon absorption by a terminated metallic nanowire}
We calculate the time dependence of the atomic excitation probability of an atom coupled optimally to the terminated metallic nanowire for a gaussian input photon wavepacket, as shown in \Fig{Terminatednanowire} (c). For such a terminated metallic nanowire, we find that the maximum of the atomic excitation probability is $72.4\%$. The fact that the probability of atomic excitation exceeds $50\%$ despite the fact that light is only injected into the system from one side, is a direct consequence of the high spontaneous emission beta factor of $91\%$ for unidirectional emission into the open end of the waveguide. For an ideal case with $\beta$-factor of $100\%$, we also find that the maximum atomic excitation is $\sim$ $80\%$ for a gaussian wavepacket. The fact that there is still a discrepancy between $72.4\%$ and $91\%$,  (resp. $80\%$ and $100\%$)   shows that gaussian input wavepackets do not form the optimum temporal pules shapes.  By pulse shaping the input wavepacket it may be possible to more perfectly approach the perfect inverting pulse.

As an example of further optimization of pulse shape, we study the possibility of shaping the input wavepacket through  waveguide dispersion. Considering a highly dispersive waveguide with dispersion relation $\alpha(\omega)=\alpha_0+\alpha_1 (\omega_0) (\omega-\omega)+\alpha_2(\omega_0)/2 (\omega-\omega)^2+\alpha_3(\omega_0)/6 (\omega-\omega)^3$, the gaussian wavepacket $\phi_i(-Z_0,t) =(2\pi\Delta^2)^{1/4}e^{[-i\omega_0 t]}e^{[-\frac{\Delta^2(-v_g t+z_0)^2}{4}]}$ initially launched at $z=-Z_0$ will be modified due to the waveguide dispersion. When the wavepacket reaches the emitter, the corresponding pulse shape is given by  \cite{Miyagi1979AO},
\begin{eqnarray}\label{DispersiPulsePropagation}
\phi_i(0,t) = & 4\sqrt{2}(\pi)^{3/4}\frac{|B|^{(-1/3)}}{\tau}e^{[(2-3AB-6C^2)\frac{1}{3B^2}-IC(3AB+2C^2-6)\frac{1}{3B^2}]}\\ \nonumber
&\times A_i[(1-AB-C^2+I2C)(|B|)^{-4/3}] ,
\end{eqnarray}
where $A=4(t-\alpha_0 Z_0)/\tau$, $B=32 Z_0 \frac{\alpha_3}{\tau^3}$, $C=8 Z_0 \frac{\alpha_2}{\tau^2}$, $\tau=4/(\Delta v_g)$, $A_i(x)$ is the Airy function.
By taking into account the dispersive features of the waveguide as well as the possibility of chirping the input pulse, we find that the chirping as well as $\alpha_2$ are simply detrimental to reaching maximum absorption, since they mainly give rise to the broadening of the pulse. Interestingly, an optimized $\alpha_3(\omega_0)$ can improve the maximum of the atomic excitation by $4\%$  compared with the nondispersive case, as shown in the contour plot  of maximized atomic excitation as function of bandwidth ($\Omega$) and $\alpha_3$ in \Fig{Terminatednanowire} (d).  In closing, we have shown that considerable amount of atomic excitation can be gained simply by operating at the termination of the plasmonic waveguide.

%

\section{Conclusion}
In conclusion, we have theoretically analyzed  the time-dependence of single-photon absorption by a single emitter. The model allows us to specify any single-photon wavepacket as initial condition, and to calculate the time dependence of the atomic excitation. We apply the theoretical model  to quasi-1D waveguides coupled to a single emitter. For a gaussian   single-photon wavepacket, we give the analytical solution to the atomic excitation, as well as the numerical results for transmitted and reflected light. We compare our time-dependent theory to the stationary theory,  and we have reported that our time dependent theory contains the stationary reflection and transmission spectra of earlier work by Shen and Fan. To optimize the excitation probability of the emitter it is essential to choose incoming wavepackets of optimum duration. Within the class of gaussian single-photon wavepackets, excitation efficiencies up to $40\%$ are possible for emitters coupled to infinite waveguides. We further studied the  impact of the finite bandwidth of the incoming wavepacket on the atomic excitation, and find the maximum excitation probablity $40\%$. This high atomic excitation simply generated by the gaussian distributed wavepacket is close to the fundamental limit of $50\%$ set by reciprocity. In order to obtain an even higher excitation probability, we have proposed to engineer the photonic environment of the emitter to suppress the spontaneous emission into all channels, except the one into which the incoming single photon is funneled.
Practically, by terminating a plasmonic nanowire and positioning the emitter properly we have found that most of the light, up to $91\%$, can be directed into can be directed into a single channel. Reciprocity guarantees that if we use the  high spontaneous emission $\beta$-factor channel for pumping, a high  atomic excitation in excess of 50\% can be achieved. Indeed, we find that a value of $72.4\%$ can be achieved by using a  very simple structure, i.e.,  a terminated metallic nanowire. This result is obtained  with a gaussian distributed single-photon wavepacket, and can be improved by shaping the optical pulse. Using waveguide dispersion for pulse shaping, a further modest improvement can be achieved.

As an outlook, we envisage that our time dependent-theory for the atomic excitation can be useful for the analysis of different experimentally relevant scenarios, and on-going experimental activities that focus on coupling a freely propagating photon to an atom \cite{Wrigge2008NaturePhy,Lindlein2007LaserPhysics,Piro2010NatPhy}. The relevance of engineering the photonic environment of the emitter is evident from our previous discussions for achieving high atomic excitation rate, and might provide new guidelines to do future experiments. Particularly promising for achieving high atomic excitation is to use  3D photonic structures, i.e., optical nanoantennas \cite{Kuhn2006PRL,Curto2010Science,Kosako2010NatPhoton}, where the temporal and spatial coherence  of the incoming wavepacket can be fully addressed. Optimizing single-photon absorption using the methods discussed in this paper will have high impact throughout the photonics community, spanning from quantum optics, single molecule absorption microscopy, to photovoltaics.


\section{Acknowledgment}
The authors wish to thank Anders S. S\o rensen, and Peter Lodahl for fruitful
discussions. We further thank Yves Rezus for careful reading of the manuscript. YTC and MW acknowledge financial support by the Danish Research Council for Technology and Production Sciences (grants no: 10$\-$093787 and 274$\-$07$\-$0080, respectively). This work is part of the research program of the ``Stichting voor Fundamenteel Onderzoek der Materie (FOM),'' which is financially supported by the ``Nederlandse Organisatie voor Wetenschappelijk Onderzoek (NWO).'' AFK was supported by a VIDI  fellowship funded by NWO.
\appendix

\appendixpage
\addappheadtotoc
\section{Normalization in continuous-mode quantization scheme}
In the well-known discrete-mode quantization scheme \cite{Loudon2000}, the commutation rules for the photon operator obey $[{\hat
a_\lambda },\hat a_{\lambda '}^\dag ] = {\delta _{\lambda \lambda'}}$. The normalization condition for these discrete modes is   $\int {\varepsilon (\bm r)[E_\lambda (\bm r)]\cdot[E_{\lambda '} (\bm r)]} dV =\delta_{\beta,\beta'}\delta_{p,p'}\delta_{m,m'}\delta_{q,q'}N_{dis}$. Accordingly, the single-photon field in the discrete-mode
quantization scheme is obtained as $\bm E_{\lambda,1,dis}(\bm r)={\bm E_{\lambda}(\bm r) \sqrt{\frac{\hbar \omega}{2N_{dis} }}}$, where $\omega=\sqrt{c^2(\beta^2+q^2)}$.
In this paper we deal with 1D waveguides, and hence a quantization box with  infinite extent parallel to the $z$ axis but with a finite cross sectional area in the transverse plane. We extend \Eq{MasterEquation} to the continuous mode quantization scheme. The mode spacing  along the $z$ direction, $\triangle\beta$, tends to zero as the quantization length along $z$ axis approaches infinity.
Therefore, the sum over all the modes $\sum\limits_{\lambda}$ can be substituted by \beq\label{SUM}
\sum\limits_{\lambda}\longrightarrow\sum\limits_{\{m, p, \beta,
q\}}\longrightarrow\frac{1}{\triangle\beta}\int
d\beta\sum\limits_{\kappa }\longrightarrow
\frac{1}{\triangle\beta}\int
d\omega\frac{1}{v_g(\omega)}\sum\limits_{\kappa }, \eeq
where $\triangle\beta$ is the mode spacings
along the $z$ direction. In this substitution rule, $\sum\limits_{\kappa }$ represents the sum over all the modes with frequency  $\omega$. The normalization
condition in the continuous-mode quantization scheme needs to be
modified to read $\int {\varepsilon (\bm r)[E_\lambda (\bm r)
]\cdot[E_{\lambda '} (\bm r)]} dV =\delta (\beta  - \beta
')\delta_{p,p'}\delta_{m,m'}\delta_{q,q'}N_{con}$. Accordingly, the single-photon field in the continuous-mode quantization scheme is obtained
as $\bm E_{\lambda,1,con}(\bm r)={\bm E_{\lambda}(\bm r)
\sqrt{\frac{\hbar \omega_\lambda}{2N_{con}}}}$. As a side remark, the discrete
Kronecker delta and the continuous Dirac delta-funtion are related by
$\delta_{\beta,\beta'}\rightarrow \triangle\beta \delta (\beta  -
\beta ')$, which indicates that the normalization factor has
incorporated a factor of $\triangle\beta$ that yields the translation of
the single-photon field  from the discrete  mode quantization to
the continuous  mode quantization as $\bm E_{\lambda,1,con}(\bm
r)=\bm E_{\lambda,1,dis}(\bm r)/\sqrt{\triangle\beta}$.
\section{Laplace transform method for gaussian input single-photon wavepacket}
An alternative approach to obtain the atomic excitation probability as a function of time that is distinct from our direct time integration of \Eq{MasterEquation1D}  to obtain \Eq{DynamicalSolution} is to use a Laplace transform method reported by Wubs \mbox{\it{et al.}} \cite{Wubs2004PRA}. In this appendix we summarize the Laplace method. By Laplace transforming the  Heisenberg's equation of motions of the atomic operator $\hat b(t)$ and field operator $\hat a(t)$, one can include the initial atomic excitations  $\hat b(t=0)$ and photonic excitations $\hat a(t=0)$ in the dynamics. More details can be found in \cite{Wubs2004PRA}. After making a pole approximation, the inverse Laplace transform gives the dynamical solution to the atomic operator,
\begin{eqnarray}\label{AtomicOperatorT}
\hat b(t)&=\hat b(0)e^{-i(\Omega_A+\epsilon)t-\Gamma t/2}\\ \nonumber
&+\frac{1}{\hbar}\int_{-\infty}^{-\infty}d\omega' [g(\omega')
\hat a_{(\omega',+)}(0)+g^{*}(\omega')
\hat a_{(\omega',-)}(0)]\frac{e^{-i\omega' t}-e^{-i
(\Omega_A+\epsilon)t-\Gamma t/2}}{\omega'-\Omega_A-\epsilon+i\Gamma/2}, \end{eqnarray}
where $\Gamma$ is the total  spontaneous-emission decay rate, $\epsilon$ is the Lamb shift of atomic transition $\Omega_A$ of the emitter due to the coupling to the waveguide modes, $g(\omega')$ is the coupling strength, $\hat b(0)$ is the initial atomic excitation,  and $\hat a_{(\omega',+)}(0)$ ($\hat a_{(\omega',+)}(0)$) is the initial right (left) propagating optical excitation. The atomic population operator $\hat N_A (t)$  is given as $\langle \hat N_A (t) \rangle=\langle i|\hat b^{\dagger}(t)\hat b(t)|i\rangle$, with the initial condition given by $|i\rangle=|g\rangle\otimes\int d\omega'S(\omega')\hat a_{(\omega',+)}(0)|0\rangle_{p}$, and $S(\omega')$ defines the incident single-photon wavepacket. Using  the same gaussian distributed single-photon wavepacket as initial condition that is also used in Sec. \textbf{3}, i.e., $ S(k)=C_k ^g(t = 0)$, one finds $
|i\rangle=|g\rangle\otimes\int dk
S(k)a_{k,+}^{\dagger}|0\rangle$,  which can be
reformulated as an integral over frequency $
|i\rangle=|g\rangle\otimes\int d\omega'
\chi(\omega')a_{\omega',+}^{\dagger}|0\rangle$, with $
\chi(\omega)=S(\frac{\omega}{v_p})/\sqrt{v_g(\omega)}$,
$a_{\omega}^{\dagger}=a_{k}^{\dagger}/\sqrt{v_g(\omega)}$.
The group velocity and phase velocity are defined as
$v_g=d\omega/dk$, $v_p=\omega/k$. Within the linear dispersion approximation, $v_g$ is equal to $v_p$. The expectation value of the excited atomic state reads as follows, \beq\label{WubsCentralResults} \langle N_A
(t)\rangle=\left|\int_{-\infty}^{\infty} d\omega
g_{\omega}\frac{e^{-i\omega t}-e^{-i
(\omega_A+\epsilon)t-(\Gamma)t/2}}{\omega-\Omega_A-\epsilon+i(\Gamma)/2}\chi(\omega)\right|^2.\eeq
Using $\sigma=\omega-\omega_0$ and omitting the Lamb shift term
$\epsilon$, one  obtains
\beq\label{WubsCentralResultsII}
 \langle N_A (t)\rangle =\left|{[\frac{2} {{\pi
{\Omega
^2}}}]^{\frac{1}{4}}}\sqrt{\frac{\Gamma}{4\pi}}\int_{-\infty}^{\infty}
d\sigma \frac{e^{[-i\sigma (t-t_0)- \frac{\sigma^2} {\Omega
^2}]}-e^{[-(\Gamma)t/2+ i\sigma{t_0} - \frac{\sigma^2}
{\Omega ^2}]}}{\sigma+i\Gamma/2}\right|^2.
\eeq
By rewriting $\frac{1}{\sigma+i\Gamma/2}$ as $\frac{1}{\sigma+i\Gamma/2}=-i\int_0^{\infty}d\xi
e^{i(\sigma+i\Gamma/2)\xi}$ and performing the $\sigma$-integral, one can further simplify $\langle N_A
(t)\rangle$ as follows,
\beq\label{WubsCentralResultsIIII}\langle N_A
(t)\rangle=|C_0^e(t)|^2 = \left|s\frac{1}{2}\sqrt{\frac{\pi }{a}}e^{\frac{b^2}{a} - c}[\mbox{\rm erf}(\sqrt a t+\frac{b}{\sqrt a})-\mbox{\rm erf}(\frac{b}{\sqrt a})]e^{- \frac{\Gamma}  {2}t}\right|^2,
\eeq
which is consistent with \Eq{DynamicalSolution}.

\end{document}